\documentclass[aps,preprint,amsmath,showpacs,nofootinbib,superscriptaddress]{revtex4}
\usepackage{array}
\usepackage{epsfig}
\usepackage{multirow}
\usepackage{stackrel}
\usepackage{latexsym,amsbsy}
\usepackage{amssymb}
\usepackage{graphicx}
\usepackage{soul}

\begin{document} 

\title{Doubly heavy baryon spectra guided by lattice QCD}
\author{H.~Garcilazo}
\email{humberto@esfm.ipn.mx}
\affiliation{Escuela Superior de F\'\i sica y Matem\'aticas,
Instituto Polit\'ecnico Nacional, Edificio 9,
07738 M\'exico D.F., Mexico}
\author{A.~Valcarce}
\email{valcarce@usal.es}
\affiliation{Departamento de F\'\i sica Fundamental e IUFFyM, Universidad de Salamanca, E-37008
Salamanca, Spain}
\author{J.~Vijande}
\email{javier.vijande@uv.es}
\affiliation{Departamento de F\'{\i}sica At\'{o}mica, Molecular y Nuclear, Universidad de Valencia (UV)
and IFIC (UV-CSIC), E-46100 Valencia, Spain}
\date{\emph{Version of }\today}

\begin{abstract}
This paper provides results for the ground state and excited spectra of 
three-flavored doubly heavy baryons, $bcn$ and $bcs$.
We take advantage of the spin-independent interaction recently obtained 
to reconcile the lattice SU(3) QCD static potential and the results of
nonperturbative lattice QCD for the triply heavy baryon spectra.
We show that the spin-dependent potential might be constrained on the basis 
of nonperturbative lattice QCD results for the spin splittings
of three-flavored doubly heavy baryons. Our results may also represent 
a challenge for future lattice QCD work,
because a smaller lattice error could help in distinguishing
between different prescriptions for the spin-dependent part of the
interaction. Thus, by comparing with the reported baryon spectra
obtained with parameters estimated from lattice QCD, one can challenge
the precision of lattice calculations. 
The present work supports a coherent description of singly, doubly 
and triply heavy baryons with the same Cornell-like interacting potential.
The possible experimental measurement of these states at 
LHCb is an incentive for this study.
\end{abstract}
\maketitle

\section{Introduction}
In a recent publication~\cite{Vij14} we have pointed
out that the static three-quark potential with parameters
determined from SU(3) lattice QCD~\cite{Tak02} does not reproduce
the triply heavy-baryon $bbb$ and $ccc$ spectra measured also in 
lattice QCD~\cite{Mei10,Mei12,Pad13}. We argued several possible 
reasons for such disagreement. In a subsequent work~\cite{Vij15}
we demonstrated that the spectra of baryons containing three
identical heavy quarks, $b$ or $c$, could be reproduced by 
means of a Cornell-like interaction, a simple Coulomb plus linear
confining potential. As it happens in the heavy meson spectra, a 
larger value of the Coulomb strength than predicted by SU(3) lattice QCD was
concluded. The phenomenological strengths of the Coulomb potential 
reproducing the heavy meson, $a$, and the triply-heavy baryon 
spectra, $A$, was found to satisfy $A/a<1/2$, slightly 
different from the $1/2$ rule as the one-gluon exchange result~\cite{Ric12}. 
The strength concluded for the linear confining interaction was also slightly
larger than the results from SU(3) lattice QCD. The spectra obtained in Ref.~\cite{Vij15}
supported a coherent description of the $bbb$ and $ccc$ heavy baryon spectra 
with the same Coulomb and confining strengths in a constituent quark model approach.

In Ref.~\cite{Vij15} it was also pointed out that the description of the $ccc$ spectra
is improved with the additional contribution of a spin-spin term, 
because relativistic effects are more important than in the $bbb$ case
and spin-dependent contributions start playing a significant role. 
Although the spin-spin interaction comes suppressed by $M_Q^{-2}$, it helped to correctly 
allocate the $ccc$ negative parity excitations with respect to the radial excitations 
of the $J^P=3/2^+$ ground state. However, due to the identity of the three quarks,
in a $QQQ$ system there cannot exist a {\it good} diquark, a couple of quarks 
with total spin 0 in a relative $S$ wave, where the spin-spin term 
is attractive and its contribution becomes relevant.

Thus, being the spin-independent part of the quark-quark interaction 
constrained by the triply heavy baryon spectra, it remains to analyze the spin-dependent 
part of the quark-quark potential. The best testing ground for this purpose are baryons 
made of three distinguishable quarks, with one of them light. 
As compared to singly heavy baryons, they are free of the uncertainties 
of the interaction between light quarks~\cite{Val08}, whose spin-dependent
part would be dominant, and the light quark kinematics~\cite{Val14}. 
Unlike triply heavy baryons, there exist pairs of quarks 
with total spin 0 in a relative $S$ wave, whose contribution will
become crucial to study the spin splittings recently reported by 
nonperturbative lattice QCD~\cite{Bro14,Pad15,Mat16}.

In this work we aim to analyze doubly heavy baryons with non-identical 
heavy quarks within a constituent quark model framework by 
means of a simple Cornell-like potential guided by lattice QCD. 
The spin-independent part of the quark-quark interaction has been determined
from the triply heavy baryon spectra recently calculated
by means of nonperturbative lattice QCD techniques and inspired
by the static potentials derived within SU(3) lattice QCD.
The spin-dependent part will be analyzed in comparison with the  
recent spin-splitting results derived in nonperturbative lattice 
QCD~\cite{Bro14,Pad15,Mat16}. 
We will show how the spin-dependent part might be constrained on the basis 
of nonperturbative lattice QCD results for three-flavored doubly heavy 
baryons, $bcn$ and $bcs$. Analogously, we will
emphasize the importance of having lattice QCD results
with smaller lattice errors, what could help in distinguishing
between different prescriptions for the spin-dependent part of the
interaction. For our purposes, we will present an exact calculation 
solving the Faddeev equations for three non-identical particles.

The road we outline in this work is similar to the path went through to
study the heavy meson spectra. Once charmonium and bottomonium spectra were 
understood within a constituent quark model framework by 
means of simple Cornell-like potentials~\cite{Qui79,Eic80},
the question of predicting and trying to understand the structure
of open-flavor mesons with a heavy-quark was soon posed~\cite{God85}.
Compared to the heavy meson case we have the great advantage of
the guidance of nonperturbative lattice QCD results and the 
static potentials derived within SU(3) lattice QCD, which 
combined with the exact method to solve the three-body problem, 
makes the difference between our work and other studies of doubly 
heavy baryons~\cite{Fle89,Kis02,Ebe97,Ton00,Ron95,Kor94,Ito00,Kau00,Ebe02,Vol01,Kar13,Yos15,Mah16,Sha16}.
Our results may also serve for a future analysis of the 
validity of the so-called superflavor symmetry, relating the spectra 
and properties of singly heavy mesons and doubly heavy baryons~\cite{Sav90,Bra05},
broken by the smallness of the heavy quark masses, which makes the size of the 
heavy diquark not small enough compared to $1/\Lambda_{QCD}$. Additional 
symmetries including excitations of the heavy diquark~\cite{Eak12} could also
be tested against our results.

A substantial basis for optimism 
in the observation of $bcn$ ($n$ stands for a light $u$ or $d$ quark) 
and $bcs$ doubly heavy baryons is the large number
of doubly heavy mesons $B_c(b\bar c)$ measured at the LHCb~\cite{Aai14}, 
indicating that simultaneous production of $b\bar b$ and $c\bar c$
pairs which are close to each other in space and in
rapidity and can coalesce to form doubly heavy hadrons is not too rare.
The cross section of pair doubly heavy diquark $(bc)$ production in high energy proton-proton
collisions has been already estimated~\cite{Tru16}.
It has also been recently discussed the
production of doubly heavy flavored hadrons in $e^+e^-$ colliders~\cite{Zhe16}
as well as the doubly heavy baryon photoproduction 
in the future $e^+e^-$ International Linear Collider (ILC) within the framework
of non-relativistic QCD~\cite{Che14}. 

The paper is organized as follows. In the next section we will briefly
review the parametrization of Cornell potential we have determined
to get a unified description of the nonperturbative 
lattice QCD $bbb$ and $ccc$ spectra. 
We will use Sec.~\ref{secIII} to discuss the solution of the 
non-relativistic Faddeev equations for three non-identical particles.
In Sec.~\ref{secIV} we will present and discuss the results 
of our work. Finally, in Sec.~\ref{secV}, we will summarize the 
main conclusions of this study. 

%
%%%%%%%%%%%%%%%%%%%%%%%%%%%%%%%%%%%%%%%%%%%%%%%%%%%%%%%%%%%%%%%%%%%%%%%%%%%
%
\section{A potential model for doubly heavy baryons}
\label{secII}

The spin-independent part of the quark-quark interaction in a baryon
should be the analog of the famous Cornell potential
for quarkonium. The short-distance behavior is expected to be 
described by the two-body Coulomb potential as the one-gluon exchange (OGE) result 
in perturbative QCD. It should be extended for the baryon case, with a 
factor $1/2$ in front of its strength due to color factors~\cite{Ric12}. 
As for the $Q\bar Q$ case, 
the characteristic signature of the long-range non-Abelian dynamics is 
believed to be a linear rising of the static interaction. Moreover, 
the general expectation for the baryonic case is that, at least 
classically, the strings meet at the 
so-called Fermat (or Torricelli) point, which has minimum
distance from the three sources ($Y-$shape configuration)~\cite{Tak04,Bor04}. 
The confining short-range potential could be also 
described as the sum of two-body potentials 
($\Delta-$shape or linear configuration)~\cite{Tak04,Bor04,Cor04,Ale03}. 
We have shown in Ref.~\cite{Vij15} the equivalence of 
both prescriptions for the case of triply heavy baryons (see 
Table II of that reference) for different values
of the heavy-quark mass.
Thus, a minimal model to study doubly heavy baryons comes given by,
\begin{equation}
V(r_{ij}) = - A \sum_{i < j}\frac{1}{|\vec{r}_i - \vec{r}_j|} \, + \, B 
\sum_{i < j}|\vec{r}_i - \vec{r}_j| \,  .
\label{Pot3Q}
\end{equation}
The value of the $Q\bar Q$ confinement strength, $b$, is usually fixed to reproduce that obtained
from the linear Regge trajectories of the pseudoscalar $\pi$ and $K$ mesons, $\sqrt{\sigma}=$
(429$\pm$2) MeV~\cite{Bal01}. In the case of baryons, the linear 
string tension $B$ is considered to be of the order of
a factor $1/2$ of the $Q\bar Q$ case. The reduction factor in the string tension can be naturally understood as a geometrical
factor rather than a color factor, due to the ratio between the minimal distance joining three quarks and the 
perimeter length of a triangle, suggesting 
$B= \left(0.50 \sim 0.58 \right) b$~\cite{Tak02}.
For the particular case of quarks in an equilateral triangle 
$B= \frac{1}{\sqrt{3}} \, b = 0.58 \, b $~\cite{Bor04}.
When the linear ansatz is adopted for the
two-body potential, still the same relation holds for the strength of the Coulomb potential $A\simeq \frac{1}{2} a$, due to color factors. 
The $\Delta$ ansatz (linear potential) has been widely adopted in the nonrelativistic quark model because of its 
simplicity~\cite{Val08,Gar07,Isg78,Oka81,Sil96,Kle10,Cre13}

On the other hand, potential models are also less accurate for baryons containing
light quarks, because relativistic effects are more important 
and spin-dependent contributions may start playing a significant role. 
Although of small importance in heavy quark systems for being suppressed
as $M_Q^{-2}$, the spin-spin interaction derived from the one-gluon exchange 
helps to improve the description of the nonperturbative lattice QCD results~\cite{Vij15}.
Thus, an spin-spin term must be considered in the interacting potential 
for those systems where spin-dependent corrections may play a role, having 
the quark-quark interaction the final form,
\begin{equation}
V_{S}(r_{ij}) = - A\sum_{i < j}\frac{1}{|\vec{r}_i - \vec{r}_j|} \, + \, B 
\sum_{i < j}|\vec{r}_i - \vec{r}_j| \, + 
A\sum_{i < j}\frac{1}{M_iM_j}\frac{e^{-r/r_0}}{rr_0^2} \left( \vec\sigma_i\cdot\vec\sigma_j \right) \,  .
\label{Pot3QS}
\end{equation}
The spin-spin interaction arising from the one-gluon exchange potential
has the same strength as the Coulomb term~\cite{Ric12}. Its
$\delta(r)$ radial structure has to be regularized in order to avoid an unbound
spectrum~\cite{Bha80}. The determination of the strength of the spin-dependent 
part in heavy baryons with three-identical quarks is not efficient.
As mentioned above, the identity of the quarks doe not allow for the existence
of a {\it good} diquark, a couple of particles with total spin 0 in a relative 
$S$ wave, where the spin-spin term is attractive and its contribution 
significant. This is why doubly heavy baryons with non-identical
heavy quarks are ideal systems to test the spin-dependent part of the
quark-quark potential. On one hand, the problem is free of the uncertainties
of chiral symmetry breaking effects related to pairs of light quarks and,
on the other hand, the distinguishability of the quarks allows for the existence of
all pairs in a relative $S$ wave with spin 0. 

Finally, to make contact with our previous studies of the heavy baryon spectra~\cite{Val08,Val14},
the linear potential is screened at long distances,
\begin{equation}
V_{SS}(r_{ij}) = - A\sum_{i < j}\frac{1}{|\vec{r}_i - \vec{r}_j|} \, + \, B' 
\sum_{i < j}\left( 1 - e^{-\mu\,|\vec{r}_i - \vec{r}_j|}\right) + 
A\sum_{i < j}\frac{1}{M_iM_j}\frac{e^{-r/r_0}}{rr_0^2} \left( \vec\sigma_i\cdot\vec\sigma_j \right) \, ,
\label{Pot3QSS}
\end{equation}
such that the same linear strength is guaranteed at short-range, $B=B' \mu$~\footnote{As
shown in Ref.~\cite{Vij04} for the light baryon spectra, the long distance screening would 
just provide with a better understanding of low-spin highly excited baryons and high-spin 
baryons, with no significant effect on the states studied on this work.}.

%
%%%%%%%%%%%%%%%%%%%%%%%%%%%%%%%%%%%%%%%%%%%%%%%%%%%%%%%%%%%%%%%%%%%%%%%%%%%%%
%
\section{Faddeev equations for three non-identical particles}
\label{secIII}

After partial-wave decomposition, the Faddeev equations are 
integral equations in two continuous variables as shown in 
Ref.~\cite{Vac05}. They can be transformed into integral
equations in a single continuous variable by expanding the two-body
$t-$matrices in terms of Legendre polynomials as shown in Eqs. (32)$-$(36)
of Ref.~\cite{Ter06}. One obtains the final set of equations,
\begin{equation}
\psi_{i;LST}^{n\ell_i\lambda_iS_iT_i}(q_i) = \sum_{j\ne i}
\sum_{m\ell_j\lambda_jS_jT_j}\int_0^\infty q_j^2 dq_j \,
K_{ij;LST}^{n\ell_i\lambda_iS_{i}T_{i}m\ell_j\lambda_jS_{j}T_{j}}(q_{i},q_{j};E)\,
\psi_{j;LST}^{m\ell_j\lambda_jS_jT_j}(q_j) \, ,
\label{e22c5}
\end{equation}
with
\begin{eqnarray}
K_{ij;LST}^{n\ell_i\lambda_iS_{i}T_{i}m\ell_j\lambda_jS_{j}T_{j}}(q_{i},q_{j};E)
&=& \frac{1}{2}
 <S_iT_i|S_jT_j>_{ST}\
\sum_{r}\tau
_{i;nr}^{\ell_iS_{i}T_{i}}(E-q_{i}^{2}/2\nu _{i})  \nonumber \\
&&\times \int_{-1}^{1}d{\rm cos}\theta \,{\frac{P_{r}(x^\prime_{i})P_{m}(x_{j})}{%
E-p_{j}^{2}/2\eta _{j}-q_{j}^{2}/2\nu _{j}}}
A_L^{\ell_i\lambda_i\ell_j\lambda_j}(p_i^\prime q_i p_j q_j)\, .
\label{e23c5}
\end{eqnarray}
$P_{r}(x^\prime_{i})$ and $P_{m}(x_{j})$ are Legendre polynomials,
$x_i^\prime=(p_i^\prime-b)/(p_i^\prime+b)$,  
$x_j=(p_j-b)/(p_j+b)$, and $b$ a scale parameter.
$\tau_{i;nr}^{\ell_iS_{i}T_{i}}(E-q_{i}^{2}/2\nu _{i})$ are the
coefficients of the expansion of the two-body $t-$matrices in terms
of Legendre polynomials defined by Eq.~(34) of Ref.~\cite{Ter06}.
$S_i$ and $T_i$ are the spin and isospin of the 
pair $jk$ while $S$ and $T$ are the total spin and isospin.
$\ell_i$ is the orbital angular momentum of the pair $jk$, $\lambda_i$ 
is the orbital angular momentum of particle $i$ with respect to the pair
$jk$, and $L$ is the total orbital angular momentum.
\begin{eqnarray}
\eta_i & = & \frac{m_j m_k}{m_j + m_k} \, , \nonumber \\
\nu_i & = & \frac{m_i(m_j+m_k)}{m_i+m_j+m_k},
\end{eqnarray}
are the usual reduced masses.
For a given set of values of $LST$ the integral equations~(\ref{e22c5})
couple the amplitudes of the different configurations
$\{\ell_i\lambda_i S_i T_i\}$.
The spin-isospin recoupling coefficients $<S_iT_i|S_jT_j>_{ST}$ are given by,
\begin{eqnarray}
<S_iT_i|S_jT_j>_{ST}  = 
(-)^{S_j+\sigma_j-S}\sqrt{(2S_i+1)(2S_j+1)} \, W(\sigma_j\sigma_kS\sigma_i;S_iS_j)
 \nonumber \\
 \times 
(-)^{T_j+\tau_j-T}\sqrt{(2T_i+1)(2T_j+1)} \, W(\tau_j\tau_kT\tau_i;T_iT_j) \, ,
\label{e8c5}
\end{eqnarray}
with $\sigma_i$ and $\tau_i$ the spin and isospin of particle $i$, and $W$
is the Racah coefficient.
The orbital angular momentum recoupling coefficients 
$A_L^{\ell_i\lambda_i\ell_j\lambda_j}(p_i^\prime q_i p_j q_j)$ are given by 
\begin{eqnarray}
A_L^{\ell_i\lambda_i\ell_j\lambda_j}(p_i^\prime q_i p_j q_j)  = 
\frac{1}{2L+1}\sum_{M m_i m_j}C^{\ell_i \lambda_i L}_{m_i,M-m_i,M}
C^{\ell_j \lambda_j L}_{m_j,M-m_j,M}\Gamma_{\ell_i m_i}\Gamma_{\lambda_i
M-m_i}\Gamma_{\ell_j m_j} \nonumber \\  \times
\Gamma_{\lambda_j M-m_j}
{\rm cos}[-M(\vec q_j,\vec q_i)-m_i(\vec q_i,{\vec p_i}^{\,\prime})
+m_j(\vec q_j,\vec p_j)] \, ,
\label{e9c5}
\end{eqnarray}
with $\Gamma_{\ell m}=0$ if $\ell -m$ is odd and
\begin{equation}
\Gamma_{\ell m}=\frac{(-)^{(\ell+m)/2}
\sqrt{(2\ell+1)(\ell+m)!(\ell-m)!}}{
2^\ell((\ell+m)/2)!((\ell-m)/2)!}\, ,
\label{e10c5}
\end{equation}
if $\ell-m$ is even. The angles 
$(\vec q_j,\vec q_i)$, $(\vec q_i,{\vec p_i}^{\,\prime})$, and
$(\vec q_j,\vec p_j)$ can be obtained in terms of the magnitudes of the 
momenta by using the relations
\begin{eqnarray}
\vec p_i^{\,\prime} &=& - \vec q_j - \frac{\eta_i}{m_k}\vec q_i \, , \nonumber \\
\vec p_j &=& \vec q_i + \frac{\eta_j}{m_k} \vec q_j \, ,
\label{e11c5}
\end{eqnarray}
where $ij$ is a cyclic pair.
The magnitude of the momenta $p_i^\prime$ and $p_j$, on the other hand,
are obtained in terms of $q_i$, $q_j$, and
${\rm cos}\theta$ using Eqs.~(\ref{e11c5}) as 
\begin{eqnarray}
p_i^\prime &=& \sqrt{q_j^2+
\left(\frac{\eta_i}{m_k}\right)^2q_i^2
+\frac{2\eta_i}{m_k}q_i q_j {\rm cos}\theta} \, , \nonumber \\
p_j &=& \sqrt{q_i^2+
\left(\frac{\eta_j}{m_k}\right)^2q_j^2
+\frac{2\eta_j}{m_k}q_i q_j {\rm cos}\theta} \, .
\end{eqnarray}

The integral equations~(\ref{e22c5}) couple the amplitude $\psi_i$ to the
amplitudes $\psi_j$ and $\psi_k$. When the three particles are different,
by substituting the equation for $\psi_i$ into the corresponding 
equations for $\psi_j$ and $\psi_k$, one obtains at best integral 
equations that involve two independent amplitudes which
means that in that case the numerical calculations are more time consuming.
If one represents in Eq.~(\ref{e22c5}) the integration over $dq_j$ by a
numerical quadrature~\cite{Abr72}, then for a given set of the 
conserved quantum numbers $L$, $S$, and $T$, 
Eq.~(\ref{e22c5}) can be written in the matrix form 
\begin{equation}
\psi_i = \sum_{j\ne i} B_{ij}(E) \psi_j \, ,
\label{mex1}
\end{equation}
where $\psi_i$ is a vector whose elements correspond to the values
of the indices $n$, $\ell_i$, $\lambda_i$, $S_i$, $T_i$, and $r$,
i.e., 
\begin{equation}
\psi_i \equiv \psi_{i;LST}^{n\ell_i\lambda_i S_i T_i}(q_r) \, ,
\label{mex2}
\end{equation}
with $q_r$ the abscissas of the integration quadrature.
The matrix $B_{ij}(E)$ is given by,
\begin{equation}
B_{ij}(E) \equiv q_s^2w_sK_{ij;LST}^{n\ell_i\lambda_i S_i T_i
m\ell_j\lambda_jS_jT_j}(q_r,q_s;E) \, ,
\label{mex3}
\end{equation} 
where the vertical direction is defined by the values
of the indices $n$, $\ell_i$, $\lambda_i$, $S_i$, $T_i$, and $r$
while the horizontal direction is defined by the values
of the indices $m$, $\ell_j$, $\lambda_j$, $S_j$, $T_j$, and $s$.
$q_s$ and $w_s$ are the abscissas and weights of the integration
quadrature.

Substituting Eq.~(\ref{mex1}) for $i=1$ into the corresponding
equations for $i=2$ and $i=3$ one obtains,
\begin{eqnarray}
\left[B_{21}(E)B_{12}(E)-1\right]\psi_2 + \left[B_{21}(E)B_{13}(E)+B_{23}(E)\right]\psi_3 = 0 \, , \nonumber  \\
\left[B_{31}(E)B_{12}(E)+B_{32}(E)\right] \psi_2 + \left[B_{31}(E)B_{13}(E)-1\right]\psi_3 = 0 \, ,
\end{eqnarray}
so that the binding energies of the system are the zeroes of
the Fredholm determinant
\begin{equation}
|M(E)|=0 \, ,
\label{mex6}
\end{equation}
where
\begin{equation}
M(E)=\begin{pmatrix}B_{21}(E)B_{12}(E)-1 & B_{21}(E)B_{13}(E)+B_{23}(E) \\
B_{31}(E)B_{12}(E)+B_{32}(E)  & B_{31}(E)B_{13}(E)-1\end{pmatrix} \, .
\label{mex7}
\end{equation}

%
%%%%%%%%%%%%%%%%%%%%%%%%%%%%%%%%%%%%%%%%%%%%%%%%%%%%%%%%%%%%%%%%%%%%%%%%%%%%%%%
%
\section{Results and discussion}
\label{secIV}
\begin{table}[t]
\caption{Faddeev amplitudes, $(\ell_i,\lambda_i, S_i,T_i)$, used in the calculation 
of the different $J^P$ states. $(L,S)$ indicates the channel giving the lowest
energy, $L$ is the total orbital angular momentum 
and $S$ is the total intrinsic spin of the three quarks. 
$\ell_i$ is the orbital angular momentum of 
the pair $jk$ and $\lambda_i$ is the orbital angular momentum of particle $i$ 
with respect to the pair $jk$. $S_i$ and $T_i$ are indicated at the bottom of the table.
For spin $1/2$ each Faddeev amplitude appears twice, with $S_i=0$ and $1$. $T_i$ is uniquely
determined, either 0 or 1/2.}
\begin{center}
\begin{tabular}{|c|p{0.5cm}cp{0.5cm}c|}
\hline
$J^P$   & & $(L,S)$   & & $(\ell_i,\lambda_i)$ \\ \hline
$1/2^+$ & &$(0,1/2)$\footnotemark[1] & & $(0,0), (1,1), (2,2), (3,3), (4,4), (5,5)$\\
$3/2^+$ & &$(0,3/2)$\footnotemark[2] & &$(0,0), (1,1), (2,2), (3,3), (4,4), (5,5)$\\																																	
$5/2^+$ & &$(2,1/2)$\footnotemark[1] & &$(0,2), (2,0), (1,1), (1,3), (3,1), (2,2), (2,3), (3,2), (3,3)$\\	
$7/2^+$ & &$(2,3/2)$\footnotemark[2] & &$(0,2), (2,0), (1,1), (1,3), (3,1), (2,2), (2,3), (3,2), (3,3)$\\	
$1/2^-$ & &$(1,1/2)$\footnotemark[1] & &$(1,0), (0,1), (1,2), (2,1), (2,3), (3,2), (3,4), (4,3), (4,5), (5,4)$\\	
$3/2^-$ & &$(1,1/2)$\footnotemark[1] & &$(1,0), (0,1), (1,2), (2,1), (2,3), (3,2), (3,4), (4,3), (4,5), (5,4)$\\
$5/2^-$ & &$(1,3/2)$\footnotemark[2] & &$(1,0), (0,1), (1,2), (2,1), (2,3), (3,2), (3,4), (4,3), (4,5), (5,4)$\\		
\hline
\end{tabular}
\footnotetext[1]{Each Faddeev amplitude may exist with $S_i=0$ and $1$ for the pair $jk$. $T_i$ is fixed, either $0$ or $1/2$.}
\footnotetext[2]{Each Faddeev amplitude is only compatible with $S_i=1$ for the pair $jk$. $T_i$ is fixed, either $0$ or $1/2$}
\end{center}
\label{tab1}
\end{table}

To obtain the predictions of the Cornell-like potential of Eq.~(\ref{Pot3QSS}) 
for the $bcn$ and $bcs$ baryon spectra to compare with the results
measured in nonperturbative lattice QCD~\cite{Bro14,Pad15,Mat16}, we solve
the three-body problem for non-identical quarks by means of the Faddeev method
described in Sec.~\ref{secIII}.
We solve the nonrelativistic Schr\"odinger equation
\begin{equation}
\left\{H_0 + V(r)\right\}\Psi(\vec r) = E \Psi({\vec r}) \, ,\nonumber
\end{equation}
where $H_0$ is the free part of quarks without center-of-mass-motion
\begin{equation}
H_0=\sum_{i=1}^3 \left( m_{i} + \frac{\vec{p}_i^{\,2}}{2m_{i}} \right) - T_{CM}\, ,\nonumber
\end{equation}
and $m_{i}$ is the mass of quark $i$. The mass of the heavy baryon will be finally 
given by $M_{B} = m_1 + m_2 + m_3 + E$. The quarks masses are taken as in 
Ref.~\cite{Val08}: $m_b=$ 5.034 GeV, $m_c=$ 1.659 GeV, $m_s= 0.545$ GeV, and
$m_u=m_d=$ 0.313 GeV, as well as the long-distance screening parameter $\mu=$ 0.7 fm$ ^{-1}$.

We show in Table~\ref{tab1} the Faddeev amplitudes that we consider to solve the 
three-body problem for each $J^P$ state, indicating the $(L,S)$ channel
giving the lowest energy. As indicated in the table, for those cases with 
intrinsic spin $1/2$, each Faddeev amplitude would contribute twice, with 
the two possible spin couplings of the $jk$ pair, 0 and 1. The isospin of the
pair $jk$ is fixed. 

Before proceeding to analyze the results, we present in Table~\ref{tab2} the convergence
of our calculation with respect to the number of the Faddeev amplitudes considered in our 
calculation, indicated in Table~\ref{tab1}. As we can see the results are fully
converged with three Faddeev amplitudes (times its degeneracy, an additional factor 
two for spin $1/2$ and another factor two for negative parity states because they
are reached with non-identical pairs of $\ell_i$ and $\lambda_i$, see
Table~\ref{tab1}), i.e., when all Faddeev
amplitudes with $\ell_i=\lambda_i \le 2$ have been considered. 
For the sake of completeness, our results have been obtained 
with all amplitudes quoted in Table~\ref{tab1}, which guarantees 
full convergence.
\begin{table}[t]
\caption{Convergence of the binding energy $E$ of different $J^P$ $bcs$ baryons, in MeV, with respect to the number, $N$,
of Faddeev configurations $(\ell_i,\lambda_i)$. See text for details.}    
\label{tab2}
\begin{center}
\begin{tabular}{|cp{0.25cm}c|cp{0.25cm}c|cp{0.25cm}c|cp{0.25cm}c|} \hline
 $N$ & & $E_{3/2^+}$ & $N$ && $E_{1/2^-}$ &  $N$ && $E_{1/2^+}$  
& $N$  && $ E_{(1/2^+)^*}$ \\ \hline
 1 &&1005 & 4  && 1226  & 2   && 953 & 2  && 987 \\
 2 && 996 & 8  && 1203  & 4   && 937 & 4  && 978 \\
 3 && 995 & 12 && 1199  & 6   && 934 & 6  && 977 \\
 4 && 995 & 16 && 1197  & 8   && 933 & 8  && 977 \\
 5 && 995 & 20 && 1196  & 10  && 932 & 10 && 976 \\
 6 && 995 &    &&       & 12  && 932 & 12 && 976 \\ \hline
\end{tabular}
\end{center}
\end{table}

\begin{figure}[t]
\vspace*{-1cm}
\mbox{\epsfxsize=110mm\epsffile{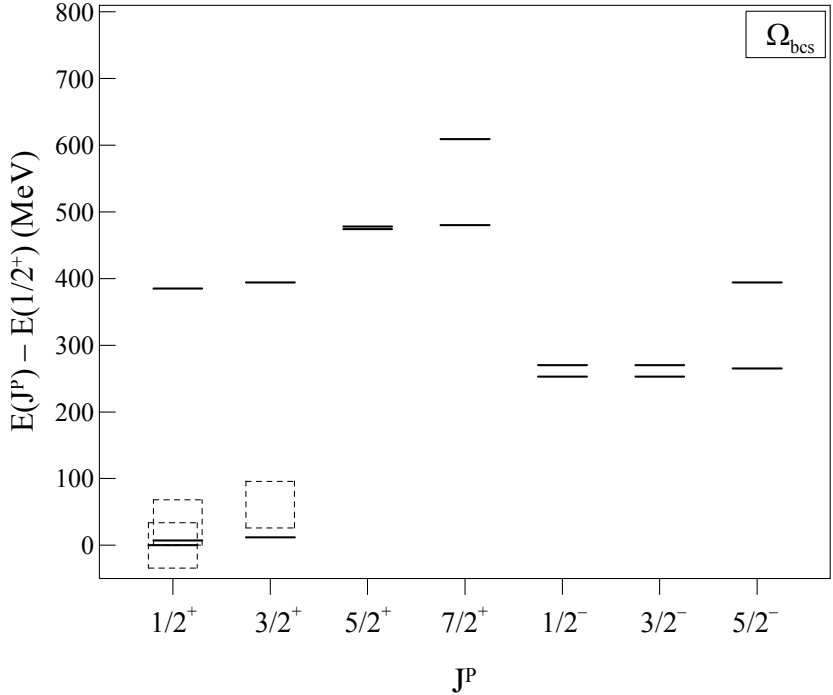}}
\vspace*{-5.5cm}
\caption{$bcs$ excited state spectra, solid lines, 
for the potential of Eq.~(\ref{Pot3QSS}) including a spin-spin interaction
regularized as in Ref.~\cite{Sil96}. The boxes stand for the
nonperturbative lattice QCD results of Ref.~\cite{Bro14}.
See text for details.}
\label{fig1}
\end{figure}

We present in Fig.~\ref{fig1} the excitation spectra of $bcs$ baryons with 
the potential of Eq.~(\ref{Pot3QSS}) and the parameters used in Ref.~\cite{Vij15}
to reproduce the nonperturbative lattice QCD results of triply heavy baryons 
$bbb$ and $ccc$: $A=$ 0.1875, $B=$ 0.1374 GeV$^2$, and an almost constant 
regularization, $r_0$, for the spin-spin term following the line of the 
model AL1 in Ref.~\cite{Sil96}.
The spin-independent part of the quark-quark interaction fixes in a unique
manner the position of the radial and orbital excitations. As explained above,
these systems provide with an additional advantage, as compared to triply heavy baryons,
that they allow to scrutinize the strength of the spin-spin interaction. In Ref.~\cite{Vij15}
it was shown how the addition of the spin-dependent part of the quark-quark
interaction, maintaining the strength of the Coulomb and the confining 
potential determined from the $bbb$ spectra, allows for a better agreement 
in the $ccc$ case. The idea behind this improvement is that
potential models probably are also less accurate for $ccc$ than 
for $bbb$ baryons, because the $ccc$ system is more relativistic 
and spin-dependent contributions may start playing a significant role. 

In a three-quark baryon, any pair of quarks must be in a $\bar {\bf 3}$ color state to
couple to a color singlet with the ${\bf 3}$ color state of the other quark. For identical
particles, the $\bar {\bf 3}$ color state is antisymmetric. Thus for a symmetric relative
$S$ wave between the quarks, $\ell_i=0$, if they are identical they can only
exist in a symmetric spin state, $S_i=1$. In other words, in a $QQQ$ triply heavy 
baryon there are no {\it good} diquarks, where the spin-dependent part of 
the interaction is attractive and significant. A pair of identical quarks could
only exist in an antisymmetric spin state, $S_i=0$, with a unit of orbital
angular momentum, which reconciles the symmetry of the state with the Pauli principle.
As the spin-spin interaction is very short-ranged, the relative $P$ wave
shields its effect. The benefit of this term in the case of triply heavy baryons
can be simply understood. In the ground state there are no pairs of quarks with 
spin zero, while there are in negative parity states. The attraction induced by
the spin-spin term allows to relax the value of the quark masses diminishing the 
repulsive effect of the centrifugal barrier. This was the main effect observed
in Ref.~\cite{Vij15}. 

By using an almost constant regularization for the spin-spin interaction,
as suggested by the model AL1 of Ref.~\cite{Sil96}, it is obtained a spin-splitting
that it is small as compared to the central value recently obtained by 
nonperturbative lattice QCD~\cite{Bro14}: $E_{\Omega'_{bc}}-E_{\Omega_{bc}}=35\pm 9 \pm 25$ MeV, and
$E_{\Omega^*_{bc}}-E_{\Omega_{bc}}=62\pm 9 \pm 25$ MeV\footnote{$\Omega_{bc}$ stands for a $bcs$ $J^P=1/2^+$ state
with two quarks in a spin 0 state, the ground state; $\Omega'_{bc}$ stands for a $bcs$ $J^P=1/2^+$ state
with two quarks in a spin 1 state; $\Omega^*_{bc}$ stands for a $bcs$ $J^P=3/2^+$ state and thus any
two quark pair is in a spin 1 state. The same notation is valid for $bcn$ states that are denoted by $\Xi_{bc}$.}. 
The predictions obtained for these spin-splittings in Fig.~\ref{fig1} are $7$ and $12$
MeV, respectively. Let us however note that these results are within 
2 sigma of the lattice central values which highlights
the importance of having smaller lattice errors to help in clearly
distinguishing between different prescriptions for the spin-dependent part
of the interaction. A similar situation is observed with the mass difference between the first two states with
$J^P=5/2^+$ or $J^P=1/2^-$, that are predicted to be almost degenerate, although in this case we have no lattice results
to compare with. This possible underestimation of the spin-spin effects 
by prescriptions as that in Ref.~\cite{Sil96} had already been noted in 
Ref.~\cite{Val08} in the study of singly heavy baryons in a constituent quark model approach, 
although, as we have mentioned in the introduction, in that case the 
presence of two light quarks did not allow a clear cut 
between the spin-independent and spin-dependent effects as in the present case, 
due to involved dynamics of the two light-quark subsystem. 
\begin{table}[t]
\caption{$E_{\Omega^*_{bc}}-E_{\Omega_{bc}}$ spin splitting, in MeV, for different prescriptions 
of the regularization of the spin-spin interaction, $r_0$. See text for details.}
\begin{center}
\begin{tabular}{|cp{0.5cm}|cp{0.5cm}cp{0.5cm}c|p{0.5cm}c|}
\hline
$E_{\Omega^*_{bc}}-E_{\Omega_{bc}}$ & & $r_{0_{(c,s)}}$         & & $r_{0_{(b,s)}}$         & & $r_{0_{(b,c)}}$   & & Latt.~\cite{Bro14} \\ \hline
12                          & & \multicolumn{5}{c|}{~\cite{Sil96}}                    & &  \multirow{5}{*}{$62\pm 9 \pm 25$} \\
63                          & & \multicolumn{5}{c|}{~\cite{Val08}}                    & &                                    \\
56                          & & ~\cite{Val08} & & ~\cite{Sil96} & & ~\cite{Sil96} & &                                    \\
22                          & & ~\cite{Sil96} & & ~\cite{Val08} & & ~\cite{Sil96} & &                                    \\
14                          & & ~\cite{Sil96} & & ~\cite{Sil96} & & ~\cite{Val08} & &                             			 \\
\hline
\end{tabular}
\end{center}
\label{tab3}
\end{table}

To illustrate the relevance of the
spin-dependent terms in three-flavored doubly heavy baryons, we evaluate the mass difference 
$E_{\Omega^*_{bc}}-E_{\Omega_{bc}}$ following different prescriptions for the regularization
of the $\delta$ term in the spin-spin interaction. The results are shown in Table~\ref{tab3}.
In the first case we use the almost constant
regularization of Ref.~\cite{Sil96} (see Eq. (2) of Ref.~\cite{Sil96}, $r_0 \in [0.26,0.37]$ fm). 
In the second case we use the flavor dependent regularization
of Ref.~\cite{Val08} (see Table 5 of Ref.~\cite{Val08}, $r_0 \in [0.017,0.27]$ fm)\footnote{A similar
recipe was used long-ago for a simultaneous study of the meson and baryon spectra in
Ref.~\cite{Ono82}.}. 
In the last three cases we identify which interaction is responsible for
the spin splitting, that as could have been expected is the spin-spin interaction between the
lightest flavors $(c,s)$, those between $(b,s)$ and $(b,c)$ being rather small.
At the light of the results of the first and the last files of Table~\ref{tab3} one can easily 
understand the results of Ref.~\cite{Vij15} regarding the spin-dependent part of
the interaction, for larger masses of the quarks the spin-splitting is almost the
same independently of the prescription used for the regularization. As explained above,
this is the reason why triply-heavy baryons are not adequate to analyze the 
spin-dependent part of the quark-quark interaction.
\begin{figure}[t]
\vspace*{-1cm}
\mbox{\epsfxsize=100mm\epsffile{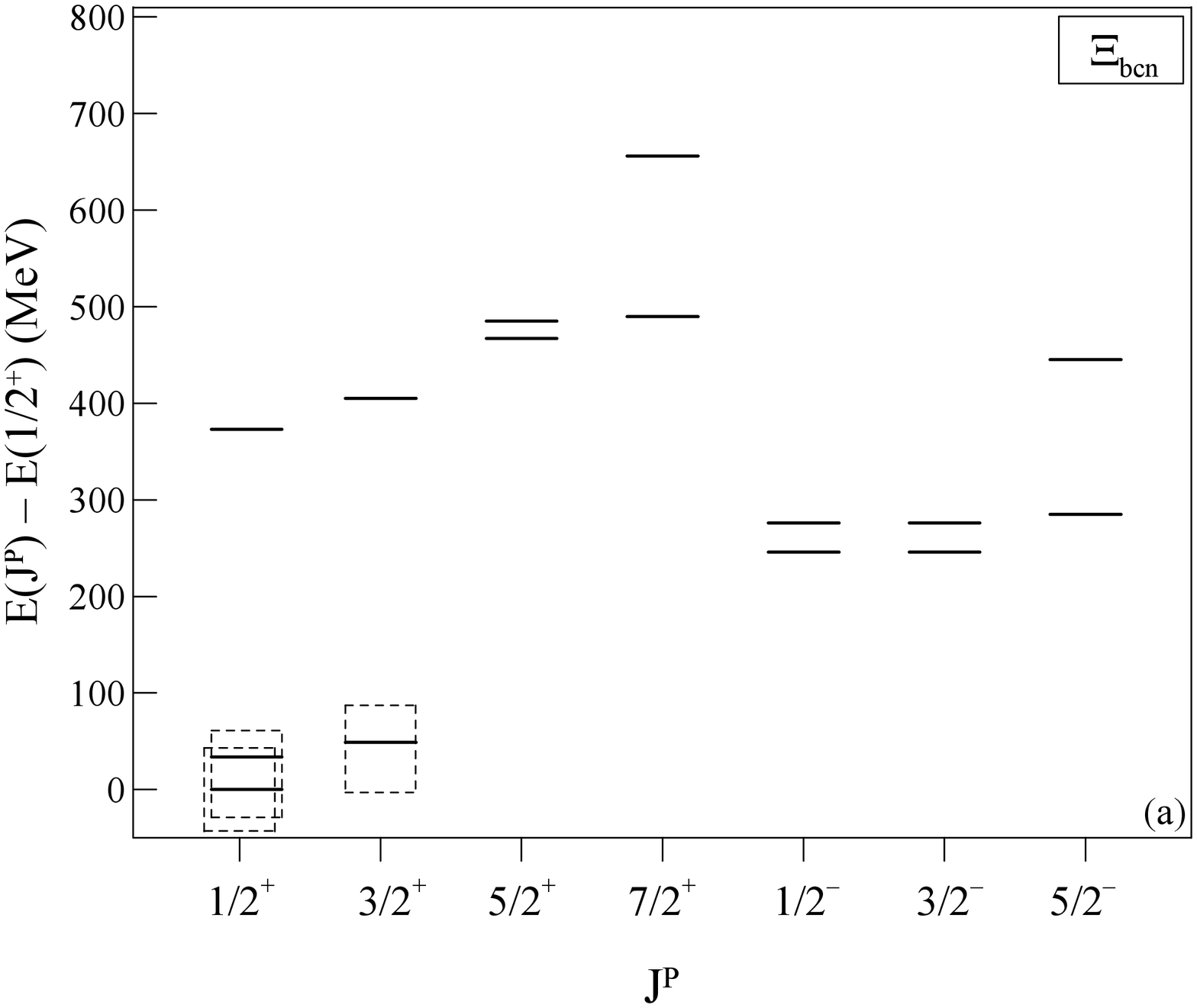}}\vspace*{-5cm}
\mbox{\epsfxsize=100mm\epsffile{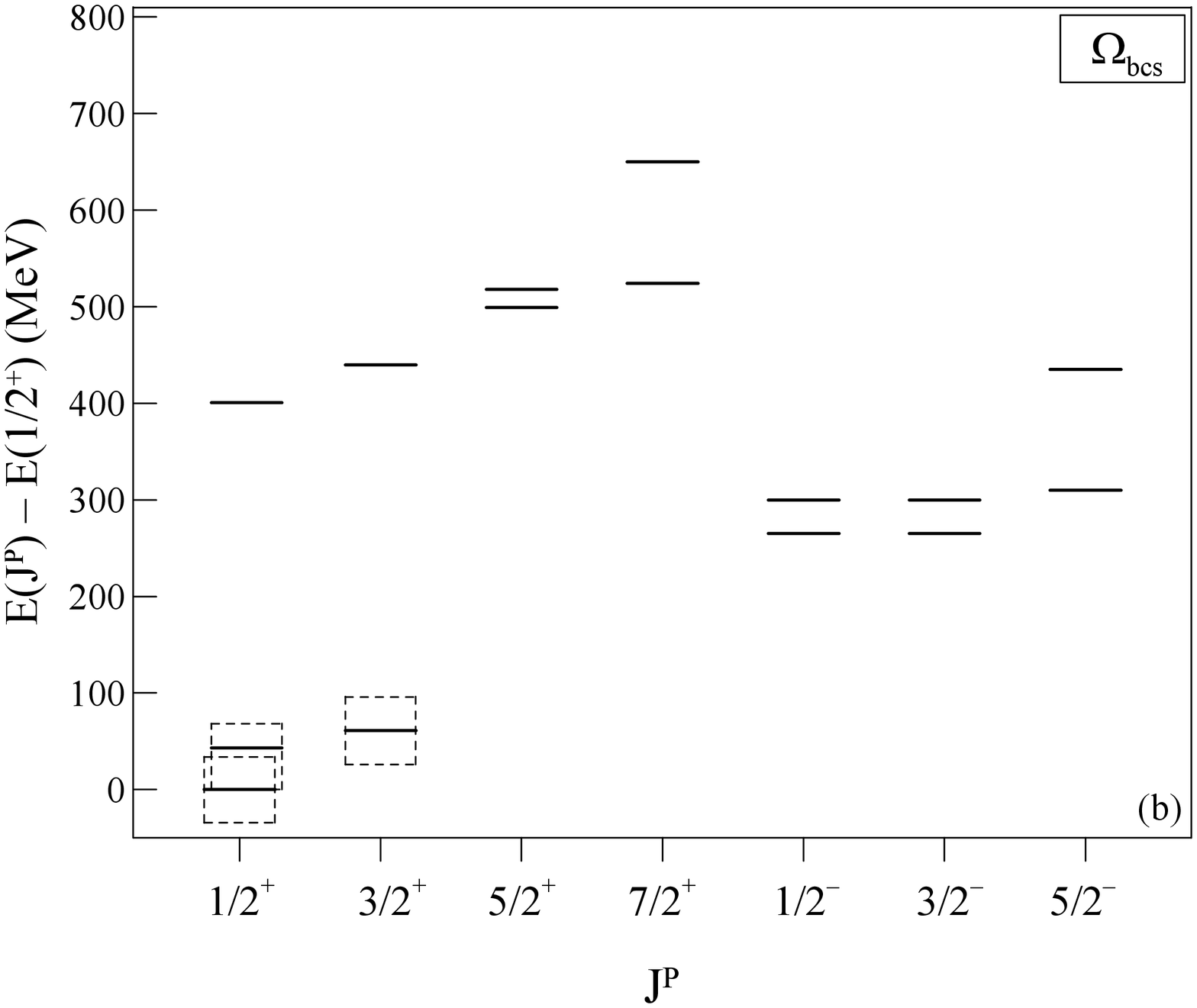}}
\vspace*{-5.5cm}
\caption{(a) $bcn$ excited state spectra, solid lines, 
for the potential of Eq.~(\ref{Pot3QSS}) including a spin-spin interaction
regularized as in Ref.~\cite{Val08}. (b) Same as (a) for
$bcs$ baryons. The boxes stand for the
nonperturbative lattice QCD results of Ref.~\cite{Bro14}. 
See text for details.} 
\label{fig2}
\end{figure}

Thus, using the flavor-dependent regularization of the spin-spin interaction derived in
Ref.~\cite{Val08} to study singly heavy baryons, we have recalculated the $bcs$ and
$bcn$ excited spectra, that are shown in Fig.~\ref{fig2}. 
As mentioned above, the spin-independent part of the quark-quark interaction 
determined in the triply heavy baryon spectra fixes in a unique
manner the position of the radial and orbital excitations, thus
these states are a first challenge for future lattice works and/or experimental
searches. Regarding the low-energy $J^P=1/2^+$ and $J^P=3/2^+$ states, the 
results in Fig.~\ref{fig2} are close to the central values of nonperturbative 
lattice QCD for the spin splitting between them~\cite{Bro14,Pad15,Mat16}. 
Thus, the present results can be considered as a useful challenge for 
future lattice QCD work because one can see how a smaller lattice error
could clearly help in distinguishing between the different prescriptions
for the spin-dependent part of the interaction.

Let us note the parameter-free nature of our calculation, making use of the 
spin-independent interaction derived in Ref.~\cite{Vij15} from the analysis of triply heavy
baryons together with the parametrization of the spin-dependent term obtained
in Ref.~\cite{Val08} from the analysis of singly heavy baryons.
Our results unify the heavy-quark dynamics for the
description of triply~\cite{Vij15}, doubly and singly~\cite{Val08} heavy baryons
by means of a simple Cornell potential with a flavor-dependent spin-spin regularization. 
They are therefore a nice testbench for future works of lattice QCD on the ground and excited
spectra of three-flavored doubly heavy baryons. They might be useful in future 
projects of lattice QCD calculations and also as a guideline
in future experiments looking for doubly heavy baryons with 
non-identical heavy quarks.

%
%%%%%%%%%%%%%%%%%%%%%%%%%%%%%%%%%%%%%%%%%%%%%%%%%%%%%%%%%%%%%%%%%%%%%%%%%%%%%%%%%%%%%%%%
%
\section{Summary}
\label{secV}

In brief, the spectra of three-flavored doubly heavy baryons
have been calculated by means of a Faddeev approach. The spin-independent
part of the quark-quark interaction was taken for grant from a 
recent study of nonperturbative lattice QCD results for triply heavy baryons. 
As in the case of the heavy meson spectra, a larger value
of the Coulomb strength than predicted by SU(3) lattice QCD is needed.
The phenomenological strengths of the Coulomb potential reproducing the heavy 
meson and the triply heavy baryon spectra satisfy $A/a<1/2$, slightly 
different from the 1/2 rule as the one-gluon exchange result. 
It has been shown that the spin-dependent part of the interaction could be fixed by 
studying the spin splitting of three-flavored doubly-heavy baryons.
The adequacy of these systems to determine the regularized $\delta$-type
interaction has been justified, obtaining a reasonable agreement with 
the central values of the spin-splittings derived by nonperturbative
lattice QCD with the same flavor dependent regularization
already used for singly heavy baryons.
Our results make evident the importance of having at our disposal 
nonperturbative lattice QCD results with smaller error,
what would allow to clearly distinguish between different 
prescriptions for the spin-dependent part of the interaction. 
Besides, they constitute a nice testbench for future works of lattice QCD on the excited
spectra of doubly heavy baryons with non-identical heavy quarks
as well as experimental searches. Let us finally note that by comparing 
with the reported baryon spectra obtained with parameters estimated from lattice 
QCD, one can challenge the precision of lattice calculations. 

The detailed theoretical investigation presented in our recent works about
the heavy baryon spectra based on nonperturbative lattice QCD guidance, may help
to improve our understanding of the interaction in many-quark systems 
containing heavy quarks, of interest to deepen our understanding 
on intriguing recent experimental results as the so-called $XYZ$ exotic 
states or the LHCb pentaquark~\cite{Che16}.
Similarly the possible advent of new experimental data~\cite{Aai14} as well as the
improvements in lattice QCD calculations of the heavy baryon spectra~\cite{Bro14},
makes the present calculation timely to scrutinize the quark-quark interaction
in systems containing heavy flavors. 

\acknowledgments
We thank to N. Mathur and S. Meinel for valuable
information about the present status of nonperturbative 
lattice QCD calculations of excited heavy 
baryon states. This work has been partially funded 
by COFAA-IPN (M\'exico), by Ministerio de Educaci\'on y Ciencia and EU FEDER under 
Contracts No. FPA2013-47443 and FPA2015-69714-REDT,
by Junta de Castilla y Le\'on under Contract No. SA041U16,
and by USAL-FAPESP grant 2015/50326-5. 
A.V. is thankful for financial support from the 
Programa Propio XIII of the University of Salamanca.

\end{document}